\documentclass[journal]{vgtc}                

\ifpdf
  \pdfoutput=1\relax                   
  \usepackage{graphicx}                
  \DeclareGraphicsExtensions{.pdf,.png,.jpg,.jpeg} 
\else
  \usepackage{graphicx}                
  \DeclareGraphicsExtensions{.eps}     
\fi%

\graphicspath{{figures/}{pictures/}{images/}{./}} 

\usepackage{microtype}                 
\PassOptionsToPackage{warn}{textcomp}  
\usepackage{textcomp}                  
\usepackage{mathptmx}                  
\usepackage{times}                     
\usepackage{cite}                      
\usepackage{tabu}                      
\usepackage{booktabs}                  
\usepackage{float}
\usepackage[usenames,dvipsnames]{xcolor}
\usepackage{caption}
\usepackage{xspace}
\usepackage{tikz}
\usepackage{graphicx}
\usepackage{csquotes}
\usepackage{wrapfig}
\usepackage{enumitem}
\usepackage{amsthm}
\usetikzlibrary{calc,shadows,shadows.blur,automata,positioning,tikzmark}

\usepackage{xspace}
\usepackage{enumitem}
\usepackage{csquotes}
\usepackage{multirow}
\usepackage{makecell}
\usepackage{listings}
\usepackage[caption=false, font=footnotesize]{subfig}
\usepackage{url}
\usepackage[scaled=.7]{beramono}
\usepackage[compact]{titlesec}
\usepackage[frozencache, newfloat]{minted}
\usepackage[font=small,textfont=up,labelfont=up,belowskip=-4pt,aboveskip=0pt,parskip=0pt]{caption}
\newtheorem{example}{Example}
\onlineid{0}

\vgtccategory{Research}

\vgtcinsertpkg

\newcommand{\sys}{\textsc{diel}\xspace}

\setminted{
  fontsize=\small,
  baselinestretch=0.7,
  fontfamily=courier,
  linenos=true,xleftmargin=\parindent,numbersep=1mm,frame=leftline}

\newcounter{hypothesiscounter}
\newcounter{subhypothesiscounter}

\usepackage[normalem]{ulem}
\definecolor{mred}{rgb}{.80,.12,.30}
\definecolor{grey}{rgb}{0.5,0.5,0.5}
\definecolor{Purple}{rgb}{.75,0,.85}

\newif\ifnotes
\notesfalse

\let\oldcite\cite
\newcommand{\ccite}[1]{\mbox{\oldcite{#1}}}

\DeclareRobustCommand{\hsout}[1]{\texorpdfstring{\renewcommand{\cite}{\ccite}\sout{#1}}{#1}}
\DeclareRobustCommand{\strike}[1]{{\ifnotes{\leavevmode\color{grey}{\protect\hsout{#1}}}\fi}}
\newcommand{\add}[1]{\ifnotes{\leavevmode\color{mred}{#1}}\else{#1}\fi}
\DeclareRobustCommand{\oreplace}[2]{\strike{#1}\renewcommand{\cite}{\oldcite}\add{#2}}
\newcommand{\replace}[2]{\ifnotes{\oreplace{#1}{#2}}\else{#2}\fi}

\ifnotes
\usepackage[textsize=small,color=lightgray,colorinlistoftodos]{todonotes}
\else
\usepackage[disable]{todonotes}
\fi

\ifnotes
\paperwidth=\dimexpr \paperwidth + 10cm\relax
\oddsidemargin=\dimexpr\oddsidemargin + 5cm\relax
\evensidemargin=\dimexpr\evensidemargin + 5cm\relax
\marginparwidth=\dimexpr \marginparwidth + 5cm\relax
\fi

\usepackage{color}

\newcommand{\engine}{\textsc{diel}\xspace}
\setminted{fontsize=\small,baselinestretch=0.7,linenos=true,xleftmargin=\parindent,numbersep=1mm,frame=leftline}
\newtheorem*{defTimespan}{Timespan}
\newtheorem*{defEvent}{Event}

\title{Programming with Timespans in Interactive Visualizations}

\author{
     Yifan Wu        \thanks{e-mail: yifanwu@berkeley.edu}  \\\scriptsize UC Berkeley
\and Remco Chang     \thanks{e-mail: remco@cs.tufts.edu}     \\\scriptsize Tufts University
\and Eugene Wu       \thanks{e-mail: ew2493@columbia.edu}    \\\scriptsize Columbia University
\and Joe Hellerstein \thanks{e-mail: hellerstein@berkeley.edu}\\\scriptsize UC Berkeley
}

\abstract{
  Modern interactive visualizations are akin to distributed systems, where user interactions, background data processing, remote requests, and streaming data read and modify the interface at the same time. This concurrency is crucial to provide an interactive user experience---forbidding it can cripple responsiveness. However, it is notoriously challenging to program distributed systems, and concurrency can easily lead to ambiguous or confusing interface behaviors. In this paper, we present \sys, a declarative programming model to help developers reason about and reconcile concurrency-related issues. 
Using \sys, developers no longer need to procedurally describe how the interface should update based on different input events, but rather declaratively specify what the state of the interface should be as queries over event history.  We show that resolving conflicts from concurrent processes in real-world interactive visualizations can be done in a few lines of \sys code.
} 

\begin{document}



\maketitle

\section{Introduction}
\label{sec:intro}

Today's interactive visualization applications are complex and difficult to develop.  
Applications need to deal with numerous concurrent processes all happening at once: user interactions in the interface; data and query processing requests to potentially large or remote databases~\cite{stolte2002polaris, liu2013immens,graphistry,mapd}; real-time data that streams into the interface\cite{xie2007towards, wanner2014state, krstajic2013visualization, zoomdata}.
Each of these processes has the capability to change contents of the user interface, and their concurrent nature can lead to modifications that confuse the user.

\begin{figure}[t]
  \centering
  \includegraphics[width=1\columnwidth]{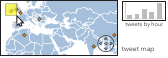}
  \caption{An interactive visualization of tweets where the tweets are streamed in real-time.}
  \label{fig:tweetVis}
\end{figure}

As a concrete example, consider the real-time Twitter analytics example in Figure~\ref{fig:tweetVis}.  The left view renders a map that is overlaid with the locations of tweets as orange dots.  New tweets stream into the visualization and dynamically add new points in the view.  The user can draw a brush on the map (the yellow box) to highlight tweets of interest.  The brush interaction interactively updates the bar chart on the right, which shows a distribution by hour of when the selected tweets were composed.  The user can click the left/right/up/down panning controls in the lower right of the map to move the map view.  Finally, the user can also click on a tweet to show the tweet's content in a pop up window.

In this example, the two processes---the user's brushing interaction, and the tweet stream---introduce ambiguity about what happens when both attempt to update the visualization concurrently.  For instance, the user intends to select the tweets on the left side of the map and is in the middle of dragging their mouse. But what if new tweets are added to the map visualization that are within the user's brush?  The user did not intend to select the new tweets, but they are now within the current brush.

\begin{figure}[t]
  \centering
  \includegraphics[width=\columnwidth]{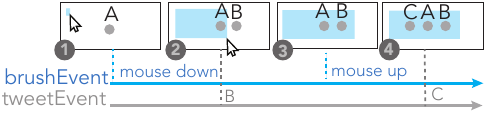}
  \caption{A frame by frame illustration of a user brushing on a map while real-time tweets are updating the UI, with time increasing left to right.  The top arrow is the timeline of brush interaction as captured by mouse events, and the lower arrow tweet update events.}
  \label{fig:realtime}
\end{figure}

\begin{example}
    Figure~\ref{fig:realtime} isolates this specific issue.  The user brushes to select point A; two new points, B and C, arrive.  We show two timelines at the bottom of the figure: the blue timeline shows mouse events related to the user's brushing
    while the gray timeline shows the new tweet arrival times.  Although it is clear that A should be selected, there are two ambiguous choices for the developer to make.  B arrived \emph{during} the brushing interaction (at step 2); should B be part of the selection?  Further, C arrived \emph{after} the interaction ended (at step 4); should C be part of the selection?  An application developer needs to choose a behavior that fits their use case, and communicate the behavior to the user.
    \label{exp:realtime}
\end{example}

The user interface updates based on events, both from the user inputs and data updates.  Events change the interface in relation to events that have happened in the past.  
Without concurrent processes, such as the brush and the tweet update, the relationships between events are simple to maintain.  For instance, for a static visualization of tweets, the events can be handled synchronously and in isolation---the brush is evaluated and the results are rendered in lockstep.  However with the added real-time tweets, the relationships between the events are made more complex.  Consider the additional event-handling logic needed to implement the policy of selecting tweet C: when handling new tweets, the developer now needs to check if the new tweet falls in the the brushed region, and if so, update the bar chart to the right to include the new tweet.
Application developers need to be aware of the concurrent processes and be able to create a \emph{policy} about how the interface should update with interleaved events from these concurrent processes.  The policies should make sense to users for their specific use case, and be easy to program.

\subsection{Hello, DIEL}

In response to these concerns, we developed \textbf{D}ata \textbf{I}nteraction \textbf{E}vent \textbf{L}og (\sys).  In \sys, the current state of the visualization is defined as queries
over the sequence of all events \add{(including user interaction events and data update events, etc.)} that have occurred in the visualization.  The queries help the developer directly access relationships between past events. The queries are expressed using familiar SQL-like syntax and semantics.
After an event, \sys stores the event to history, and updates the new state of the visualization based on the queries specified by the developer.
\sys manages the state of the application in the face of concurrent processes and does not replace the entire application---it integrates within existing applications through a JavaScript interface.

For example, Listing~\ref{lst:realtime} shows the example of how to express a policy \add{for the brushing example in Figure 2}.
Lines 1-4 of Listing~\ref{lst:realtime} define the history of events in \sys, specifying what the attributes are for each type of event---the brush event contains the bounding box as well as what the current mouse state is (down, move, or up), and the tweet event contains the id of the tweet, the geo location associated with the tweet, and the time it was tweeted.  The application can use simple APIs  to inject the events into \sys.

Upon the arrival of any event (either a new brush event \emph{or} new tweet event), the \texttt{regionSelection} view in lines 6-10 of Listing~\ref{lst:realtime} provide SQL-like \sys queries over those tables.  It defines the tweets  currently selected by the brush by computing a join between the tweet data events and the latest brush interaction (\texttt{LATEST brushEvent}).  The join identifies tweets whose lat, lon coordinates are within the bounding box of the current brush (\texttt{is\_within\_box()}).
\add{Because of the use of the LATEST brush coordinates, all points A, B, and C will be selected since all points fall within the bounding box of the brush (and the timing of the events is not considered).}
Other policies, such as to only select A, or to cancel the brushing interaction if a new tweet arrives that changes the set of selected tweets, are also easy to implement in \sys, as we show in Section~\ref{sec:patterns}.

\begin{listing}[ht]
\begin{minted}{sql}
CREATE INPUT brushEvent(mouseEvent TEXT,
  latMin REAL, lonMin REAL, latMax REAL, lonMax REAL);
CREATE INPUT tweetEvent(tId TEXT, content TEXT,
  lat REAL, lon REAL, hour INT);

-- region brush: include A, B, and C
CREATE VIEW regionSelection AS
  SELECT * FROM tweetEvent AS t
  JOIN LATEST brushEvent AS currentBrush
  WHERE is_within_box(t.lat, t.lon, currentBrush);

CREATE OUTPUT hourDistOutput AS
  SELECT hour, COUNT() FROM regionSelection GROUP BY hour;
\end{minted}
  \caption{Example code snippet that define one policy for the concurrent brush and tweet.
  The \sys keyword \texttt{INPUT} specifies events. \texttt{LATEST} identifies the most recent events. Function \texttt{is\_within\_box} checks if the lat and lon fall within the bounding box.}
  \label{lst:realtime}
\end{listing}

\add{The last two lines (12-13) of Listing~\ref{lst:realtime} provide the query to retrieve data for visualizing the barchart in Figure 1 where information about the points A, B, and C (which are selected by the brush) are shown.}

\begin{figure}[t]
  \centering
  \includegraphics[width=1\columnwidth]{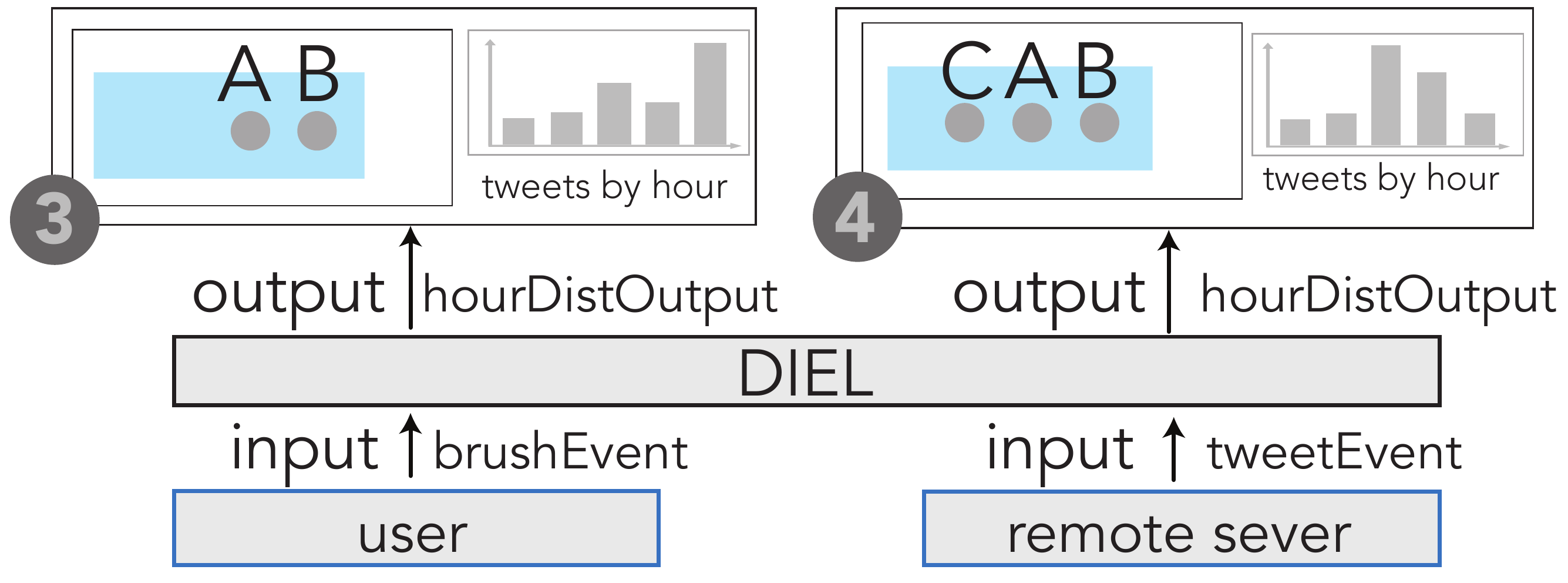}
  \caption{
  A closeup of events 3 and 4 from Figure~\ref{fig:realtime}.  The \sys engine first takes a \texttt{brushEvent} from the user, and then a \texttt{tweetEvent}, which produces visualization states 2 and 3, respectively.  Internally, \sys advanced a logical timestep for every new event, appends them in their corresponding history logs, and updates the state by executing the query-based policies.  Thus, the visualization state is well defined after every event.}

  \label{fig:api}
\end{figure}

Given this specification, \sys will ensure that the interface will automatically include the new tweet when it arrives and falls into the current brush, along with any other state of the visualization that may depend on it; in this case the bar chart to the right.  This is illustrated in Figure~\ref{fig:api}, a simplified diagram of the \sys execution process.

To summarize, we contribute by (1) identifying a class of challenges caused by concurrent timespans in interactive visualizations, and (2) developing a framework, \sys, for state management in interactive visualizations that simplifies handling of concurrent timespans.  
We demonstrate that \sys facilitates simple, and we believe elegant programming of the underlying state of interactive visualizations that deal with concurrent events, while interoperating with a wide range of toolkits for visualizing data.  We also show that a wide range of other functionalities for visualizations, such as undo/replay~\cite{shneiderman2003eyes}, logging, and interaction summaries~\cite{feng2017hindsight}, can be naturally supported by \sys.
\section{Timespans in Interactive Visualizations}
\label{sec:diagnosis}

We begin by providing a conceptual model to think about concurrent processes during interactive visualization, which helps clarify the problem space.  We use the language of \emph{events} and \emph{timespan}s to discuss the challenge of concurrent processes:

\begin{defEvent}
  An event is any change from outside the application that is captured by the application, such as a user click, or a data response from server.
\end{defEvent}

\begin{defTimespan}
  Events form a linear history that can be numbered sequentially to define a logical notion of time, and a timespan is a tuple composed of a time of a starting event, and time of an ending event.
\end{defTimespan}

\begin{figure}[t]
  \centering
  \includegraphics[width=0.7\columnwidth]{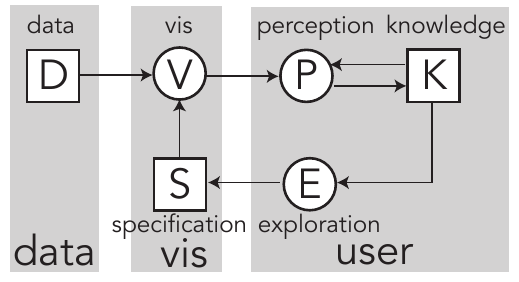}
  \caption{A model of interactive visualization. Circles are processes and rectangles are inputs and outputs.}
  \label{fig:vismodel}
\end{figure}

We now provide example categories of timespans and the corresponding events using a model of visualization by van Wijk~\cite{van2005value}, reproduced in Figure~\ref{fig:vismodel}.  The application uses \emph{data} and a visualization \emph{specification} to define a visualization.  This is perceived by the user and adds to the user's \emph{knowledge}.  Based on this new knowledge, the user then explores new specifications, which starts the process again.  By examining different arrows in the model, we can characterize common instances of timespans.
Broadly, events and timespans can be classified as either from user interactions or from data updates, and timespans may contain a mixture of different event types.

Timespans are necessarily developer-defined, as their definition are inherently determined by the application's semantics.
The taxonomy provides developers with a structure to identify relevant timespans for their application.

\subsubsection{Data Timespans}

\begin{itemize}[leftmargin=*]
  \item \emph{Data \& Specification$\rightarrow$Vis: data processing timespan}. Applications may need time to process the specification when the data is large or lives on a remote server.  To get from data to the final image, the application may incur network transmission, server processing, and perhaps rendering delays.  Different types of events may be part of this timespan, including the interaction that initiated the request, the processing request, the response, and rendering events.
  \item \emph{Data$\rightarrow$Vis: data streaming timespan}: real-time data are generated over time (e.g., tweet streams), which creates a timespan that starts from the start of the visualization user's session to the end of the session.
\end{itemize}

\subsubsection{User Timespans}

\begin{itemize}[leftmargin=*]
  \item \emph{Perception$\rightarrow$Exploration: reaction and think timespans}: users may take time from observing a change on the screen to taking action~\cite{card1986model, kosinski2008literature, reactiontime}.
  \item \emph{Exploration$\rightarrow$Specification: user action timespan}: some interactions, such as drawing a brush, or making multiple selections by clicking on each item, take time to perform the physical action.
  \item \emph{Specification$\rightarrow$Vis: user selection timespan}: often selections persist on the screen after the event is completed, such as a brush.  This selection is perceived by the viewer and may still impact other events (e.g., new tweets, or a map resize).  The start of the selection timespan is the creation of the selection until the removal of the selection.
\end{itemize}

\noindent The above is not intended to be an exhaustive list, but serves to showcase the rich varieties of timespans that may be present in modern, highly interactive visualizations.  These different timespans are often due to concurrent processes (e.g., server processing, users, network, etc), which we show is a major source of concurrency ambiguities below.

\subsection{Overlapping Timespans in the Wild}

The ubiquity of timespans in modern visualizations makes overlaps between timespans highly likely.  
We now present examples from each of the three classes of timespan overlaps: user-user, user-data, and data-data.



\subsubsection{User-User Timespan Overlap}

The following is an example of the overlap due to two user interactions.  
\begin{example}
  Some user interactions can have persistent effects to the visualization. For example, suppose that the user can draw a persistent brush (i.e. a brush that stays on the screen to indicate a user's interest in monitoring a specific region in the map). Then the user pans the map.
  These two user selection timespans overlap each other because the brush's timespan continues ``infinitely'' into the future.  This results in ambiguities in how the system should respond.
  Should the brush continue to persist?  If so, and user's intent was to monitor a geographic region, then the brush could essentially move off-screen after the pan.  But how would they know that it is still active and affecting the visualizations linked?  Otherwise, should the brush stay in the same pixel region, and select different geographic regions during the pan?
  \label{exp:useruser}
\end{example}

\subsubsection{User-Data Timespan Overlap}

User timespans may overlap with data timespans, and are common because the system runs concurrently to the user.   In addition to the example in the introduction, we now describe two additional examples.

\begin{example}
  The user perceives an interesting tweet, A, and proceeds to click on it. However, 10 milliseconds before they were able to click, the map data updates and the mouse position the user clicked on corresponds now to tweet B.  The application now updates the charts with the selection of tweet B.  However, we know that no human user can observe and act on a visual change in 10 milliseconds~\cite{card1986model}, so the selection of B must have been unintentional.  This type of overlap is common across selection widgets where the contents may change due to fine-grained user interaction, such as drop-down menus for auto-completion or search suggestions\add{, or a user accidentally clicking on a banner ad in a webpage due to asynchronous loading using AJAX}.
  \label{exp:reaction}
\end{example}

\strike{This overlap introduces conflict between a user's intention and that captured by the application.  Without treatment, the interface executes a command that the user did not intend.  The user may not even have noticed the error, in which case they get wrong results.  Or if the user has noticed the error, it would either take effort to undo the interaction, or it may not even be possible (e.g., if they irrevocably dispatched the wrong task).}

Often in interactive visualizations, there are implicit \emph{dependencies} between events.  For instance, if the user is drawing a brush on a map, the brush depends on the position of the map to remain stable.
Below is an example of such dependency potentially being violated due to overlapping timespans.

\begin{example}
  Suppose that instead of streaming tweets, the tweets are being fetched from the server for each new map position.  The map view does not change position until the data has been loaded (data processing timespan).
  A user first clicks the right arrow to navigate to a new map region, but for a short while, this map does not change.  During this time, the user notices an interesting tweet, and clicks on it to see the content. 
  However immediately after, the data for the new map region (from the first interaction) arrives.  Should the application block the new map region?  Or should it have blocked the click on the tweet?
  \label{exp:dataproc}
\end{example}

\subsubsection{Data-Data Timespan Overlap}

In traditional interfaces that execute synchronously, data processing follows lock step after each interaction. However, with a client-server architecture that introduces data-processing timespans, this will no longer be the case; instead, limits should be posed on what sequence of events are acceptable, depending on the application.

\begin{example}
  Consider an interactive cross-filter interface over bar charts (e.g., ~\cite{xfilter}). As the user brushes a line chart, fine-grained requests are being sent to fetch the filtered data for each new range selected.
  Ideally, the sequence of request and response should be in lock step, \texttt{(q1} \texttt{r1} \texttt{q2} \texttt{r2} \texttt{q3} \texttt{r3)}
  where \texttt{q} is the request and \texttt{r} is the response.
  However, these requests experience varying response delay and instead the following event sequence happens \texttt{(q1} \texttt{q2} \texttt{r1} \texttt{q3} \texttt{r3} \texttt{r2)}.
  For each of the events, what response should the application render? Should it render \texttt{r1} \texttt{r3}, then \texttt{r2}? Or should it ignore \texttt{r1} and \texttt{r2} and just render \texttt{r3}?
  \label{exp:preempt}
\end{example}

\section{Reasoning About Concurrency}
\label{sec:rational}

While each of the examples above can be resolved using any popular programming paradigm for user interface design, we will argue, in this section, that we have a simpler approach for managing timespans in interactive visualizations.  The crux of our argument is that by keeping a log or history of all events, timespans can easily be specified as queries over that history, and these queries can be extended to specify how overlapping timespans should be treated.
We begin by explaining how it is not possible to simplify interface programming by preventing concurrency.  

\subsection{The High Cost of Preventing Concurrency}

\begin{figure}[h]
  \centering
  \includegraphics[width=0.7\columnwidth]{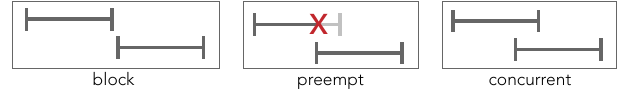}
  \caption{Possible strategies for dealing with overlapping timespans. The lines segment represents a single timespan (defined by the start and end), with time going from left to right.  The first two strategies prevent the overlap by forcing the start of one timespan to align with the end of another.}
  \label{fig:timespans}
\end{figure}

Overlaps can clearly cause problems, and it is not simple to reason about how to handle overlaps effectively---they resemble concurrency issues in distributed systems, which are notoriously complex.
As a result, a common design approach is to preclude such overlaps entirely.
Prevention can be achieved by forcing each new event to wait, and starting only after the previous event has completed (blocking).  Alternatively, it can be achieved by forcing the previous timespan to complete early when the later timespan begins (preemption).  We illustrate the two choices in Figure~\ref{fig:timespans}.  Unfortunately, both options can have high costs, as we now show.

The first option, blocking, is benign when the backend and network are fast enough that each timespan completes before another or is perceived to start and when there are no real-time events.
This is often not the case and improving performance may present a significant challenge.
One way to avoid backend delays is to load all the data needed in advance, so that interactions can proceed synchronously with negligible latency.  This is common practice in commercial charting programs like Tableau~\cite{tableau}.  However, pre-loading all data on the client complicates the application logic and may not be possible when the data set is large.

The second option, programmatic preemption, is common among frameworks for front-end application development (e.g., ~\cite{elmHttp}).  While this may seem like a simple solution, it may not always apply.  For instance, in Example~\ref{exp:dataproc}, the event handlers do not \emph{know} that there is a dependency between the brush and the map; to program this dependency, the developer would have to coordinate the handling of the map and brush interactions.  Preemption may also not be the best design.  For instance in Example~\ref{exp:preempt}, the user may want to see intermediate results that are only a few milliseconds late.

Given that overlap prevention is not a panacea, how should a developer reason about and program concurrent events?  
To account for the overlap between timespans, we need to handle each event based on past events.  To coordinate with past events, a developer can either (1) derive and maintain sufficient information needed from past events, or (2) store all past events and examine past events for the information needed.  The former programs \emph{without} history, and the latter \emph{with} history.
We discuss these two alternatives and highlight pain points to motivate the \sys model.

\subsection{Programming Without History}

To maintain relevant information from past events, a developer can create and manage variables in their event handlers, and use these variables to control how events are handled.

We walk through how to deal with the example in Figure~\ref{fig:realtime}.
Under traditional event handling, A and B are selected, since a call back re-evaluates the selection criteria each time the brush moves.
In order to select all of A, B, and C, the developer first needs to save the bounding box of the brush, and filter every new Tweet (e.g., C) to check if it falls within the bounding box.  If so, then the new Tweet is added to a separate variable that stores the selected Tweets.  This logic then needs to be disabled once the brush is cancelled. 
Now if the developer want to change it such that only A is selected, since it is the only tweet that was in the region before the brush has happened, the developer need to mark the new tweets as having started after the brush has started, and when evaluating the brush, skip those marked.

Without history, the programmer needs to perform imperative bookkeeping, and as a result when the design changes, they need to modify different parts of the event handling and bookkeeping logic.  In most UI packages, the state captured is scattered across variables in multiple handlers, which further exacerbates the challenge of derived state management.

  
\subsection{Programming With History}

When programming with history, the developer defines declaratively, \emph{what} the next state of the UI is, without needing to procedurally describe \emph{how} to get there from the current state. The relevant state of the UI can be captured as data; the graphical elements can be fully derived from that data, via a  graphical grammar~\cite{wilkinson2006grammar}.  Interactions can also be seen as additional data to be visualized, e.g., the state of a brush on a map at any point in time is defined by two coordinates (\texttt{latMin}, \texttt{lonMin}, \texttt{latMax},\texttt{lonMax}), and map as a semi-transparent rectangle to the screen.

As an example, Listing~\ref{lst:realtime} simply defines the state of the UI as queries over history---the developer no longer needs to track the current state of the variables and UI, nor do they need to define the different ways that one state could change to the next.




There are many ways to represent event history, and even more ways to program with it.  We chose to program history with SQL because it is (a) a ``declarative'' or ``functional'' approach where the output is computed at any time from a spec over history, (b) a simple language with modest temporal power---just enough to support simple overlap checks among timestamps and timespans, and the ability to query history with simple predicates like ``most recent event matching a condition'', (c) a language that is familiar to our target audience, namely developers of data visualizations, and (d) widely implemented on browsers, mobile devices, and backend servers; this accelerated our development of \sys.

We cannot claim in any empirical sense that the language in \sys is the perfect choice, but we believe it is a good fit to our design considerations, and as a result a good fit to taming the complexity of overlapping timespans.

\section{The DIEL Model}
\label{sec:model}
We now introduce the \sys model: how different data in the interactive visualization is modeled (data model), and how application state evolves (execution model).

\subsection{Data Model}

Data in \sys are represented as relational tables with rows and columns. There are four types of data in any application: input, history, static, and output; we discuss each type of table in turn.

\subsubsection{Input Tables} contain events from the world that may change the application state.  The events may be from the user (e.g., clicks), or asynchronous data inputs (e.g., new tweets).  Inputs are unpredictable, and the application (and \sys) has no control over when they might arrive. Within a \sys program, input tables are read-only.  They are populated by the application through the \sys API.  

\subsubsection{History Tables} are used by a \sys developer to persist data derived from input events, or storing data that does not to be updated with each event.  They are INSERT-only tables that the developer can define and populate in the \sys program.  
Deletions and updates are not allowed---we keep history immutable to make reasoning about program behavior easier in the face of concurrency~\cite{helland2015immutability, coblenz2016exploring, immutablejs}.  The \texttt{allBrushes} table in Listing~\ref{lst:undo} is an example for supporting UNDO functionality.

\subsubsection{Static Tables} store pre-populated data that is READ-only within the \sys program.  For instance, the map tiles in the Tweet visualization is fixed and does not change.   Since \sys is evaluated using a database, static tables are simply existing tables in the database.

\subsubsection{Output Views} represent the state that the application ultimately reads and renders in the visualization.  The purpose of the \sys program is to automatically maintain these views by executing queries over the input, history, and static tables.   For instance, \texttt{hourDistOutput} in Listing~\ref{lst:tweet} defines data used to render the bar chart in Figure~\ref{fig:tweetVis}.


\subsection{Execution Model}

The developer defines the above tables, and writes {\it state programs} that specify how the history tables are populated when different types of events arrive.  Developers can also define database views to organize the overall \sys program to help define the Output Views.  A database view is a table whose rows are not explicitly stored in the database but are computed as needed from a query~\cite{ramakrishnan2000database}.
Once these are defined, \sys automatically performs the following sequence of actions to update the Output Views on every new event that arrives in the Input Tables:

\begin{enumerate}[leftmargin=*]
  \item Increment a global logical timestep, and annotate event with logical timestep and physical wall-clock time.
  \item Store input event in corresponding Input Table.
  \item Execute the corresponding \emph{state programs} to populate history tables.
  \item Execute/maintain the Output Views
  \item Invoke upcalls to the visualization for each Output View that has changed. 
\end{enumerate}

\section{The DIEL Prototype}
\label{sec:implementation}

To evaluate the feasibility and expressivity of \sys, we implemented a prototype that runs in a standard web browser.  The \sys language is SQL with some syntactic sugar and and \sys program compiles to a JavaScript library that a developer can reference. The library runs on a browser-based SQLite engine~\cite{sqljs}
This section introduces the syntax through a naive implementation of the tweet example.  We walk through syntax for setting up the inputs, outputs and how to use the compiled JavaScript object.  
In the next section, we will show how to use \sys to resolve overlapping timespans.  For the ease of reading, we will follow the convention of naming inputs with suffix \texttt{Event}, and outputs with \texttt{Output}.

\subsection{Input/Output Syntax}
Listing~\ref{lst:realtime} lines 1-4 shows how to create input tables.
The \sys keyword \texttt{CREATE INPUT} takes after \texttt{CREATE TABLE} syntax in SQL, with the specification for the column names, their data types, and additional constraints.  Listing~\ref{lst:tweet} lines 1-3 show the other two events: map panning (\texttt{mapEvent})  and click to select an individual Tweet (\texttt{clickEvent}).

Listing~\ref{lst:tweet} lines 5-7 show the syntax for creating output views.
\texttt{CREATE OUTPUT} specifies a query over \sys tables.  \texttt{mapOutput} queries for the most recent map panning event. The \texttt{LATEST} keyword is syntactic sugar to find the records with the highest logical timestep.  The query for \texttt{mapOutput} is the same as \texttt{SELECT *} \texttt{FROM} \texttt{mapEvent} \texttt{WHERE} \texttt{timestep} \texttt{=(SELECT} \texttt{MAX(timestep)} \texttt{FROM mapEvent}.

\begin{listing}[ht]
\begin{minted}{sql}
CREATE INPUT mapEvent (
  latMin REAL, lonMax REAL, latMin REAL, lonMax REAL);
CREATE INPUT clickEvent (lat REAL, lon REAL);

CREATE OUTPUT mapOutput AS SELECT * FROM LATEST mapEvent;
CREATE OUTPUT brushOutput AS SELECT * FROM LATEST brushEvent;
CREATE OUTPUT tweetsOutput AS SELECT * FROM regionSelection;
\end{minted}
  \caption{Additional input and output definition for the tweet visualization besides those provided in Listing~\ref{lst:realtime}.}
  \label{lst:tweet}
\end{listing}

\subsection{API Syntax}

Given a \sys program, \sys generates a Javascript object \texttt{diel} the application can import and use.
The input specifications are translated into API function calls of the form 
 \texttt{diel.input.<inputName>}.  We show examples of how the application can invoke it within its user interaction event handlers (Listing~\ref{lst:apiInputItx}) and for streaming data (Listing~\ref{lst:apiInputData}).

\begin{listing}[ht]
\begin{minted}{js}
brush = d3.brush() // brush interaction handlers
  .on("move", function() {
    // .. map extent to lat/lon bounding boxes ..
    diel.input.brushEvent({mouse: 'move',
      latMin, latMax, lonMin, lonMax});
  })
  .on("end", function() { 
  	// .. map extent to lat/lon bonuding boxes ..
    diel.input.brushEvent({mouse: 'end',
    latMin, latMax, lonMin, lonMax});
  }); 
\end{minted}
  \caption{Input user interaction event in the tweet streaming example, where the events are captured by a D3 brush.  Brush extents need to be translated into a lat/lon bounding box, which are then passed to \sys.}
  \label{lst:apiInputItx}
\end{listing}

\begin{listing}[ht]
\begin{minted}{js}
// data input from websocket
socket.onmessage = function (event) {
  diel.input.tweetEvent(event.data);
}
\end{minted}
  \caption{Input data event in the tweet streaming example, where tweets are streamed in from a WebSocket connection.}
  \label{lst:apiInputData}
\end{listing}

The output views are complied into API calls of the form \texttt{diel.bindOutput.}\texttt{<outputName>}.  The application passes in a callback function responsible for taking the full set of records in the updated Output View and rendering it on in the UI (Listing~\ref{lst:apiOutput}).  We expect that the function is state-less and idempotent, meaning that calling it multiple times with the same records results in the same output rendering.   Supporting deltas and incremental updates is an optimization left for future work. 

\begin{listing}[ht]
\begin{minted}{js}
function update_bar_chart(data) {
  changeSet = embed.vega.changeset()
    .remove(embed.vega.truthy)
    .insert(data)
  view.change('table', changeSet).run()
}
diel.bindOutput.hourChartOutput(update_bar_chart);
\end{minted}
  \caption{Example use of \sys output API.  \texttt{update\_bar\_chart()} uses Vega~\cite{satyanarayan2017vega} to render the bar chart in Figure~\ref{fig:tweetVis}.  It is called by \sys whenever the \texttt{hourChart} output view changes.}
  \label{lst:apiOutput}
\end{listing}

\subsection{State Program and History Tables Syntax}

The syntax of a state program is \texttt{CREATE PROGRAM} \texttt{BEGIN} \texttt{<program} \texttt{Content>} \texttt{END;}.  For instance, in Listing~\ref{lst:history}, the history table is \texttt{selections} \texttt{History} and the program content is the \texttt{INSERT} clause, which saves the selection of tweets that was executed at each timestep.
If the state program should run after specific events, the developer can use the following syntax: \texttt{CREATE PROGRAM} \texttt{AFTER} \texttt{<inputName>}.


\begin{listing}[ht]
\begin{minted}{sql}
CREATE TABLE selectionsHistory AS (tId TEXT);
CREATE PROGRAM -- run after all interaction events
  BEGIN -- store the selected tweets, represented by tId
    INSERT INTO selectionsHistory
      SELECT tId from regionSelection;
    -- developers can write more queries here
  END;
\end{minted}
  \caption{Illustration of using a state program to append to history tables.  At every timestep, the selection made is persisted into the history table, \texttt{selectionsHistory}, if there are no selections, there will be no records.  \sys automatically keeps track of the timestep for all history tables.
  }
  \label{lst:history}
\end{listing}



\section{Resolving Timespans With DIEL}
\label{sec:patterns}

This section will use \sys to address the examples provided in Section~\ref{sec:intro} and Section~\ref{sec:diagnosis}.
We show how we can change the overlap policies easily, often with just a few lines of code.  The goal of our discussion is not to prescribe specific design policies, but rather to facilitate implementation.

\subsection{User-User Timespan Overlap}

Recall Example~\ref{exp:useruser}, where the brush bounding box persists and  the user pans the map.  We show  two possible ways to resolve the ambiguity. Recall that the \texttt{brushOutput} table defines the state of the user's brush.  We can specify new policies by redefining it based on the brush's relationship with the panning input table.   

Alternative 1 in Listing~\ref{lst:clearBrush} implements a policy that 
 ``removes'' the brush after a map interaction by redefining the brush state to be \texttt{brushPreemptedOutput}, which will be empty if a map interaction exists after the last brush interaction event.    Alternative 2 implements \texttt{brushStillInOutput}, which is empty if the latest brush's bounding box is no longer within the latest view box as defined by the panning event.  


\begin{listing}[ht]
\begin{minted}{sql}
CREATE OUTPUT brushPreemptedOutput AS -- alternative 1
  SELECT * FROM LATEST brushEvent AS b
    JOIN LATEST mapEvent AS m ON b.timestep > m.timestep;

CREATE OUTPUT brushStillInOutput AS -- alternative 2
  SELECT * FROM LATEST brushEvent AS b
    JOIN LATEST mapEvent AS m ON is_within_box(b, m);
\end{minted}
  \caption{Example \sys code to implement two different conflict resolutions: \texttt{brushPreemptedOutput} ``removes'' the brush if there is a newer map interaction, and \texttt{brushStillInOutput} ``removes'' the brush if it is no longer visible in the current map view.}
  \label{lst:clearBrush}
\end{listing}

\strike{However, the map could move to a new position where the brush is no longer visible; because the brush is still filtering the distribution by hour chart, it may be confusing to the user what the bar chart data is.
One fix is to ``remove'' the brush selection if the map interaction is more recent, which is implemented in output \texttt{brushPreemptedOutput} Listing~\ref{lst:clearBrush}, which declares that the brush is only defined if it is more recent than that last map interaction.}

\strike{After some discussion with users, the designer for the project decides to try  another policy where the brush selection should only be reset if it is not fully visible in the map.  For instance, when the map zooms out, the brush selection stays the same, but if the map moves to a new region where the brush is no longer contained, the brush selection should be empty.  This can be easily implemented, as shown in output \texttt{brushStillInOutput}, by filtering on the brush's bounding box in relation to that of the map's.}

\subsection{User-Data Timespan Overlaps}


\subsubsection{User Action and Data Streaming Overlap}

Listing~\ref{lst:realtime} demonstrated one possible policy to resolve the conflict in Example~\ref{exp:realtime} by selecting all points (A, B, C).    An alternative policy is to only  select the tweets that were on the screen at the time the brush began (i.e. point A).  Thus, neither B nor C should be part of the selection. In Listing \ref{lst:selectAlt}, \texttt{initialSelection} (lines 2-5) expresses this by finding the most recent mouse down event from \texttt{brushEvent} and ensuring that only the tweets in the bounding box (\texttt{regionSelection}) that also have a lower timestamp are kept.

Alternatively, if the designer chooses to invalidate a brush when new tweets arrive, they can do so by selecting the most recent brush only there does not exist a panning interaction that is more recent (Listing \ref{lst:selectAlt}, lines 9-14.)

\begin{listing}[ht]
\begin{minted}{sql}
-- include only A
CREATE VIEW initialSelection AS
SELECT * FROM regionSelection AS r WHERE r.timestep < 
  (SELECT timestep FROM LATEST brushEvent
    WHERE mouseEvent='down');

-- return most recent brush contents
-- unless a tweet arrived after most recent mouseDown
CREATE VIEW filteredBrush
  SELECT * FROM LATEST brushEvent
  WHERE NOT EXIST (
    SELECT * FROM tweetEvent WHERE timestep > (
      SELECT timestep FROM LATEST brushEvent
        WHERE mouseEvent='down')
  );
\end{minted}
  \caption{Two different policies for selection over real-time data.  \texttt{initialSelection} selects only the tweets that are present at the time of the initial brush, and \texttt{filteredBrush} removes the brush if there has been a new tweet since the brush started.}
  \label{lst:selectAlt}
\end{listing}

For completeness, event deletions are also possible, albeit slightly more cumbersome, when using this immutable history paradigm.   This can be achieved by keeping track of events as either additions or deletions (\texttt{tweetEvent} in Listing~\ref{lst:removeTweet}), and writing a query that ignores events that have a later deletion event (\texttt{tweetOutput}). 


\begin{listing}[ht]
\begin{minted}{sql}
CREATE INPUT tweetEvent (
  lat REAL, lon REAL, tId TEXT,
  opType TEXT check (opType = 'add' OR opType = 'del')
);

CREATE OUTPUT tweetOutput AS
  SELECT * FROM tweetEvent t
  WHERE t.opType = 'add';
    AND NOT EXISTs ( -- filter out the deleted records
      SELECT 1 FROM tweetEvent s
      WHERE s.lat = d.lat AND s.lon = d.lon
        AND s.uId = d.uId AND s.opType = 'del');
\end{minted}
  \caption{An immutable implementation of streaming tweet data where tweets can be removed.  For each added tweet, it check that a later deletion event doesn't exist.}
  \label{lst:removeTweet}
\end{listing}

\subsubsection{User Reaction and Data Streaming Overlap}

We now discuss Example~\ref{exp:reaction}, where the user inadvertently clicks on the wrong tweet because it was inserted immediately before the user's click action.  Listing~\ref{lst:reaction} shows two possible policies to address this issue.

\texttt{skipUnintendedClick} checks that the time of the user's click event did not occur within 200ms~\cite{kosinski2008literature} of the most recent tweet event in the \sys system.  If so, the click event is simply discarded.  The 200ms can and should be adjusted based on the use case and users.

\texttt{intendedSelect} illustrates a second policy that simply ignores tweets that were introduced within 200ms of the user's mouse click.  This is done by identifying the tweet in the clicked location with a timestamp that is least 200ms earlier than the click event's timestamp.

\begin{listing}[t]
\begin{minted}{sql}
CREATE VIEW skipUnintendedClick AS 
  SELECT *  
  FROM latest clickEvent AS currentClick
  WHERE timestamp > (
    SELECT max(timestamp) FROM tweetEvent
    WHERE currentClick.lat = tweetEvent.lat
      AND currentClick.lon = tweetEvent.lon
  ) + 200; -- 200 ms approximates human reaction time

CREATE VIEW intendedSelect AS
  SELECT * FROM LATEST clickEvent AS currentClick
  JOIN tweetEvent
    ON tweetEvent.timestamp < currentClick.timestamp - 200
    AND currentClick.lat = tweetEvent.lat
    AND currentClick.lon = tweetEvent.lon;
\end{minted}
  \caption{Two different policies for dealing with user reaction timespan.  The first \texttt{skipUnintendedClick} filters out interrupted brush interactions. and the second \texttt{intendedSelect} selects based on the state which is buffered by reaction time.
  }
  \label{lst:reaction}
\end{listing}


\subsection{Data-Data Timespan Overlap}

Example~\ref{exp:preempt} discusses what happens when the brush sends asynchronous requests and the responses arrive out of order.  One possible overlap policy is to display the response as soon as it arrives if it is more recent that that shown previously.  For example, \add{assume an intended sequence of requests \texttt{[1, 2, 3, 4, 5]}.} If the \replace{response were from interactions}{the received order of the sequence is} \texttt{[1, 4, 3, 5, 2]}, then \add{following this policy,} the rendered result will be \texttt{[1, 4, 4, 5, 5]} for each timestep.
This shows partial results that are slightly out of sync, but also prevents the potential issue of showing an incorrect final result, as well as out of order updates.   

\strike{This is implemented in Listing~\ref{lst:newerOnly}}\add{Listing~\ref{lst:newerOnly} shows an implementation of this policy}.  The brush response input event keeps track of what the timestep of the initial request as the attribute \texttt{requestTimestep}.  \texttt{carrier} and \texttt{flightCount} are the actual response payloads.   \texttt{barChartOutput} picks the response of the most recent request by sorting the responses in descending order of \texttt{requestTimestep}.  


\begin{listing}[ht]
\begin{minted}{sql}
-- using the flights cross-filter example
CREATE INPUT brushResponseEvent (
  requestTimestep INT, carrier INT, flightCount INT);

CREATE OUTPUT barChartOutput AS
  SELECT * FROM brushResponseEvent
  ORDER BY requestTimestep DESC LIMIT 1;
\end{minted}
  \caption{An implementation of showing results that are from newer interactions.}
  \label{lst:newerOnly}
\end{listing}

To summarize, all the design policies in the examples can be expressed by straightforward selection and join queries that are easier to write and understand than state that is manipulated and stored in variables throughout the application logic.  This allows the developer to quickly iterate on different concurrency policies to find those most effective for their application.
\section{From Fixing Conflicts to Creating Features}

The previous sections demonstrated how \sys can be used to help developers resolve different types of timespan conflicts. However, the power of \sys is not limited to this capability.  In fact, the key concept behind \sys -- the use of event history to reason about  visualization state -- can be extended to implement other capabilities. 

Shneiderman has argued for recording a history of actions to support undo, replay, and progressive refinement~\cite{shneiderman2003eyes}.
Since \sys records history, these important features are easy to support.
Below we present how \sys can be used to support undo, provide history-based scents, perform logging and debugging, and facilitate information sharing in a client-server architecture.

\subsubsection{Undo}

History is immutable, thus \sys models an undo as another interaction event.  We will show linear undo, used in applications such as Emacs~\cite{prakash1994framework}, in the Tweet example.  For simplicity, we assume that the user click on Tweets to select them.  She clicks Tweets A, B, and C, and then presses undo twice.  The sequence of actions is (A, B, C, Undo, Undo), but what the user will see is the sequence of {\it selections}: (A, B, C, \underline{B}, \underline{C}).  where \underline{B} mean that B is selected due to an undo action.

Listing~\ref{lst:undo} provides an implementation.  We first define an undo input table; \texttt{allSelections} will track the sequence of selections as described above, where the rank is simply the sequential id.   The state program populates \texttt{allSelections} with the actual selected tweet for each logical timestep (i.e., each click and undo event).  To compute the selected tweet, it first checks whether the most recent interaction were undo events or normal clickEvents (lines 23-25).  If there are undo events, then it subtracts their count from the maximum rank in \texttt{allSelections} to identify the selection that the undos represent (e.g., \underline{B} after the first undo above).  Otherwise, \texttt{currentUndoSel} is empty.    

\texttt{currentSelection} uses the SQL \texttt{coalesce} function to either pick the selection from the \texttt{currentUndoSel}, or if it is empty, to pick the most recent brush event.  Finally, the state program adds the current selection.

\begin{listing}[ht]
\begin{minted}{sql}
CREATE INPUT undoEvent(); -- undo interaction, with no parameters

CREATE TABLE allSelections(tweet int,
  rank int autoincrement);

CREATE PROGRAM -- record the items at every timestep into state
  BEGIN INSERT INTO allSelections 
        SELECT * FROM currentSelection; END;

CREATE VIEW currentSelection AS
  -- `coalesce` picks the first non-null argument.
  SELECT coalesce(currentUndoSel.tweet, clickEvent.tweet) 
  FROM LATEST clickEvent LEFT OUTER JOIN currentUndoSel ON true;

CREATE VIEW currentUndoSel AS
  SELECT * FROM allSelections AS s
  WHERE rank = ( -- past rank as the result of undo operations
    SELECT max(rank) FROM allSelections -- the current timestep
  ) - ( 
    SELECT count() * 2 - 1 -- the count of new undos minus 1
    FROM undoEvent u
    WHERE u.timestep > (select max(timestep) FROM clickEvent)
  );
\end{minted}
  \caption{\engine code for simply implementation of linear undo.}
  \label{lst:undo}
\end{listing}

\subsubsection{History-Based Scents}
\begin{figure}[t]
   \includegraphics[width=0.4\linewidth]{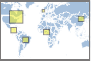}
    \begin{minted}{sql}
CREATE OUTPUT brushScentOutput AS
  SELECT DISTINCT * FROM brushEvent WHERE mouseEvent='up';
    \end{minted}
  \caption{Example facet widget in the style of \emph{HindSight}}
  \label{fig:hindsight}
\end{figure}

The \emph{HindSight} design technique represents interaction history directly in the data visualizations to enhance the user experience~\cite{feng2017hindsight}. Figure~\ref{fig:hindsight} illustrates an example for the tweet visualization, where the regions that the user has brushed are visualized in a world map view widget.  The visualization is that of the final selection (filter by mouse up), and only selects distinct values (a convenient SQL functionality).

\subsubsection{Logging and Debugging}

Since all the events are logged, developers can save the \sys tables from user sessions to the server, which can be queried at a later time, to answer questions such as, how many interactions were made, and what the average latency was per requests.  The authors have used this functionality to replay past user sessions as part of debugging the visualization~\cite{narayanasamy2005bugnet, reduxTools, elmreplay}.

\subsubsection{Client-Server Architectures}

Accessing remote servers directly is another functionality that is supported by the \sys model. Listing~\ref{lst:remoteDirect} shows a join between a \sys table \texttt{mapEvent} and a table \texttt{remote.tweets} that is stored on a remote database server. Here, the new input \texttt{tweetEvents} is defined as a query over a remote data source based on the user's panning interactions.  This allows the request to be evaluated asynchronously, and be handled simply as another \sys input table.    This functionality is not yet implemented, because it requires a federated SQL layer~\cite{sheth1990federated} that executes across multiple databases.   However, the techniques are well understood, and we plan to borrow techniques from existing federated systems such as BigDAWG~\cite{Gadepally2016TheBA}.

\begin{listing}[ht]
\begin{minted}{sql}
-- query remote databases directly
CREATE INPUT tweetEvent AS
  SELECT tId, lat, lon, hour
  FROM remote.tweets AS t JOIN LATEST mapEvent AS currentMap
    ON is_within_box(t.lat, t.lon, currentMap)
  WHERE retweetCount > 10000;
\end{minted}
  \caption{Using \sys to directly query remote data sources.  Defining the query as an input table automatically makes the query asynchronous.}
  \label{lst:remoteDirect}
\end{listing}



Another common functionality for applications with client-server architectures is to upload the current state of a client application to the server, useful for migrating a session from one device to another~\cite{ghiani2010demand} or for restoring a user session after a network disconnection.  To implement the feature, the developer shares \sys tables on the client to servers, and the complete session can be directly reconstructed base on \sys tables.
\section{Related Work}
\label{sec:related}

Our work presents the challenge of \emph{timespans} in interactive visualization, which broadly falls in the category of ``concurrency'' issues, which has been explored in much depth in other fields such as databases and collaborative groupware.  While the root cause of the problems are similar in that events do not occur in lockstep, the challenges they cause are very different.  In this section, we show how timespans differ from prior models of concurrency/asynchrony, and how \sys uniquely addresses the issues of timespans.

\subsection{Concurrency in Databases}

Databases developed the formalism of transactions as the foundation of concurrent execution of reads and writes to the tables.  The properties that a transaction must maintain is that the transactions are atomic---all the reads/writes are executed in full, or none are and that the transactions are isolated from other concurrent transactions.  While \sys makes use of the relational representation, it does \emph{not} use the transaction formalism to reason about timespans.  In \sys there is only one thread running, appending one event at a time. When \sys discusses concurrency, it's the concurrency between timespans, not between insertions.  For future work, we could investigate how to map timespans into transactions. However that will be work building on top of \sys, not instead of it.

There is no notion of \emph{locking} in \sys's context since there are no concurrent record access. Note that this should not be confused with \emph{blocking} in the UI; blocking is handled in \sys not by rejecting input record but by not outputing the most recent event.

\subsection{Operational Transforms}

Collaborative editors also face the issue of ``concurrent'' operations across users.  However, interactive visualizations  express queries over data for a single user, and groupware express modifications of shared objects for applications like shared documents and multi-player games.

The focus in collaborative editing is on distributed \emph{consistency}: the issues that arise when multiple views (replicas of system state) are subject to viewing and modification, and each replica may learn about updates to system state in different orders resulting in potential inconsistency~\cite{ellis1989concurrency,sun1998operational}.  \sys instead addresses the case where multiple processes (e.g., user interactions, responses to previous asynchronous requests, streaming data) modify a single visualization state. Thus, all events are strictly ordered and recorded in the log, and the \sys system is able to see the timespans formed by pairs of ordered events. This is arguably a simpler problem than the above (which has a reputation for being complex~\cite{sun1998operational} and difficult to prove). 

\sys has a single event log and does not need to keep replicas of the application state in sync, which is much simpler. \sys provides a programming abstract to enable programmers to design the correct behavior easier, and not an algorithm that imposes a design choice.  That being said, integrating the two approaches for \emph{collaborative visualizations} is interesting future work.

\subsection{Programming Abstractions for Concurrency and Asynchrony}

Outside of application verticals, there has been several domain specific languages designed to deal with concurrency, ranging from functional reactive programming (FRP), temporal logic, to constraint programming.  We discuss each in turn.

In \emph{Reactive Extensions}, events are modeled as streams that can be composed with functional operators such as \texttt{switch} and \texttt{reduce}~\cite{meijer2012your}.
Reactive Vega is another popular functional reactive programming framework~\cite{satyanarayan2016reactive}, with a focus on interactive visualizations.  Reactive Vega generates a dataflow graph based on declarative specifications and can efficiently process events.  

Both systems are fundamentally streaming (``push'') driven systems, which encourage a reactive coding style similar to state machines, where state transitions and emits output as input flows through. However it is feasible to use these FRP languages as a substrate to incrementally maintain both a history and the result of standing queries over that history: this is how stream queries and relational materialized view maintenance work. In short, one could probably implement \sys-like logic as an idiosyncratic design pattern in an FRP language like Rx or Vega, but doing so would not be a task for a basic visualization developer.

\emph{Bloom} provides a simple temporal model, consisting only of timestamped data, event-driven timesteps, and relational queries run at each timestep~\cite{alvaro2011consistency}.  \sys applies principles from the Bloom language to interactive applications, and adapts a SQL interface as opposed to Datalog, which is lesser known and presents a deeper learning curve.

\sys is also related to the Forward~\cite{fu2014forward} system and the DeVIL language~\cite{wu2016devil,wu2017combining}.   Both are database abstractions for application programming.
Forward is a ``full stack'' framework where developers can directly access remote tables.  Forward differs from \sys in that it synchronizes the client and server state and does not provide abstractions for asynchronous events.  DeVIL is similar to \sys in that it defines declarative SQL-like abstractions to express interactions and how they change the visualization state.  However, it does not explicitly store event history, nor let developers query the history.

\emph{Constraint programming}~\cite{jaffar1994constraint,oney2012constraintjs}  automatically maintain developer specified constraints over variables, and allow the constraints to be context-dependent to the input sequence via FSM abstractions.
\sys does not allow developers to reason about state transitions for individual events. Although our language is more restrictive, it is sufficiently expressive for programming concurrent events in interactive visualizations---as opposed to generic interfaces.


\sys expose the log as a simple query-able table, and make it easy to express and directly see the effects of a wide variety of reconciliation policies within the visualization, in order to choose the policy that provides good user experience.

\section{Conclusion and Future Work}
\label{sec:conclusion}

In this paper, we present DIEL, a model that allows a developer to reason about timespans and write programs for resolving conflicts from overlapping timespans. 
We implemented a prototype DSL and demonstrate the ease with which DIEL supports different designs for a variety of real-world examples.
We also showed that the history \sys keeps facilitates functionalities such as undo, replay, and logging.

Future work directions include further system improvements, as well as exploration of the design space and policies for overlapping timespans. In the near term we plan to open source \sys and release a gallery of examples to get feedback from the interactive visualization community. On the system front, we plan to enhance the performance of the \sys engine by exploring (a) storage reclamation for unreachable history~\cite{conway2014edelweiss}, (b) avoidance of recomputation via materialized view maintenance techniques~\cite{chirkova2012materialized} and lineage~\cite{psallidas2018smoke}, and (c) optimized federated query execution across remote and local data~\cite{sheth1990federated}. On the policy level, we are exploring the design space for timespans involving asynchronous remote calls, performing user studies to better understand how design affordances can enable users to exploit and keep track of the multitasking that is made possible by well-managed overlapping timespans. Over time, we hope to generalize to higher-level principles for design recommendations for timespans, much as the work on grammars of graphics have led to design recommendations for static visualizations~\cite{mackinlay1986automating,mackinlay2007show,wongsuphasawat2017voyager}.




\section{Acknowledgments}
This work was supported by the National Science Foundation under Grant No. 1564351, 1527765, and 1564049.

\bibliographystyle{abbrv-doi}

\bibliography{ref}
\end{document}